# Chemical Reaction Engineering and Catalysis:
## AI/ML Workflows and Self-Driving Laboratories

Rigoberto Advincula [1, 2]* and Jihua Chen [1]

[1]Center for Nanophase Materials Sciences, Oak Ridge National Laboratory (ORNL)
1 Bethel Valley Road, Oak Ridge, TN 37830

[2]Department of Chemical and Biomolecular Engineering,
University of Tennessee at Knoxville
1512 Middle Dr, Knoxville, TN 37996

This manuscript has been authored by UT-Battelle, LLC, under Contract No. DEAC05-00OR22725 with the U.S. Department of Energy. The United States Government and the publisher, by accepting the article for publication, acknowledge that the United States Government retains a nonexclusive, paid-up, irrevocable, worldwide license to publish or reproduce the published form of this manuscript, or allow others to do so, for United States Government purposes. DOE will provide public access to these results of federally sponsored research in accordance with the DOE Public Access Plan
(http://energy.gov/downloads/doe-public-access-plan).

* To whom correspondence should be addressed: radvincu@utk.edu

## ABSTRACT

Chemical reaction engineering is key to industrial might and sustainable chemistry. This will be enabled using smart, efficient catalysts or catalysis ecosystems. This is possible with advanced artificial intelligence and machine learning (AI/ML) workflows that need to be employed as agentic AI projects. The fundamentals of catalysis need to be emphasized. A strong focus on catalyst design, mechanistic studies, reaction engineering, and scale-up must use ML-driven workflows, along with high-throughput experimentation (HTE) and an autonomous, self-driving laboratory (SDL). Laboratory experience and data-driven approaches are valuable when working together to accelerate this development. Parametrize and create a virtuous circle for data-driven discovery across heterogeneous, homogeneous, and biocatalysts to enable utility in many chemical process industries as agentic AI tasks. This article builds the case for discovery science in catalysis and continuous improvement in chemical reaction engineering with this new ecosystem.

# 1. Introduction

Chemistry is the central science intersecting with many branches of science to explain the phenomena of matter.[1] Its practicality lies in the ability to explain observable phenomena through quantifiable observations using the scientific method to test theories. The reality is grounded in the principles of thermodynamics, quantum mechanics, and spatio-temporal scale, as explained by the branch of science called physical chemistry. Applied chemistry (which used to exist as a field or department in some universities) or chemical technology can deliver real-world solutions and industrial might but must be operationalized with engineering—a field or profession that is called chemical engineering.

Chemical engineering is a branch of engineering that applies chemistry, physics, math, and biology to design large-scale processes that transform raw materials (such as fossil fuels, minerals, biomass, or even living cells) into valuable and practical products.[2] These can range from fuel, plastics, food, feeds, pharmaceuticals, and critical minerals. The engineering focus can be on safety, efficiency, sustainability, and economics. The ability to convert raw and basic materials into other chemical intermediates or products requires process control. This process can involve chemical, physical, or biological conversion or transformation, which can also be quantified per unit time or at a given scale (space, volume, etc.) and involves control of pressure (P), temperature (T), volume (V), solubility, viscosity, etc.[2] This is where advances in discovery chemistry and mechanistic studies can be missed in translational research. It is important to view this phase as "breaking the mold" where the stage is not just process optimization but discovery science. [3]

Another important task in chemical engineering is scale-up from the bench to the chemical plant.[3,4] The upstream and downstream need to be connected in a supply chain that accounts for the conversion process, ensuring it is both economically and thermodynamically efficient, safe, and sustainable. The task, then, is to take lab-scale-developed reactions and translate them into actual industrial processes. It must be techno-economically feasible, circular in its life cycle, and designed for process safety. This is how chemical and biomolecular manufacturing advances, leading to new investments and markets. [5]

## 2.Reaction Engineering

Chemical reaction engineering starts with sound chemistry. Mechanistically, the energetics of the reaction, the transition state, reaction rate laws, etc., can be modeled at the atomic or quantum level.[5] This defines the transformation in terms of bond energies, which can be audited at each step, and includes free-energy, enthalpy, and entropic considerations. With the chemistry established, reaction engineering can now be parametrized by external factors that control equilibrium conditions or shift the system toward product formation and its efficiency under controlled PVT conditions. It is sometimes called the process intensification stage (Figure 1). [6] Mixing, flow, and viscoelastic behavior, or the lack of their optimization, are key to reaction engineering and design of manufacturing systems, from reactors to separation units.

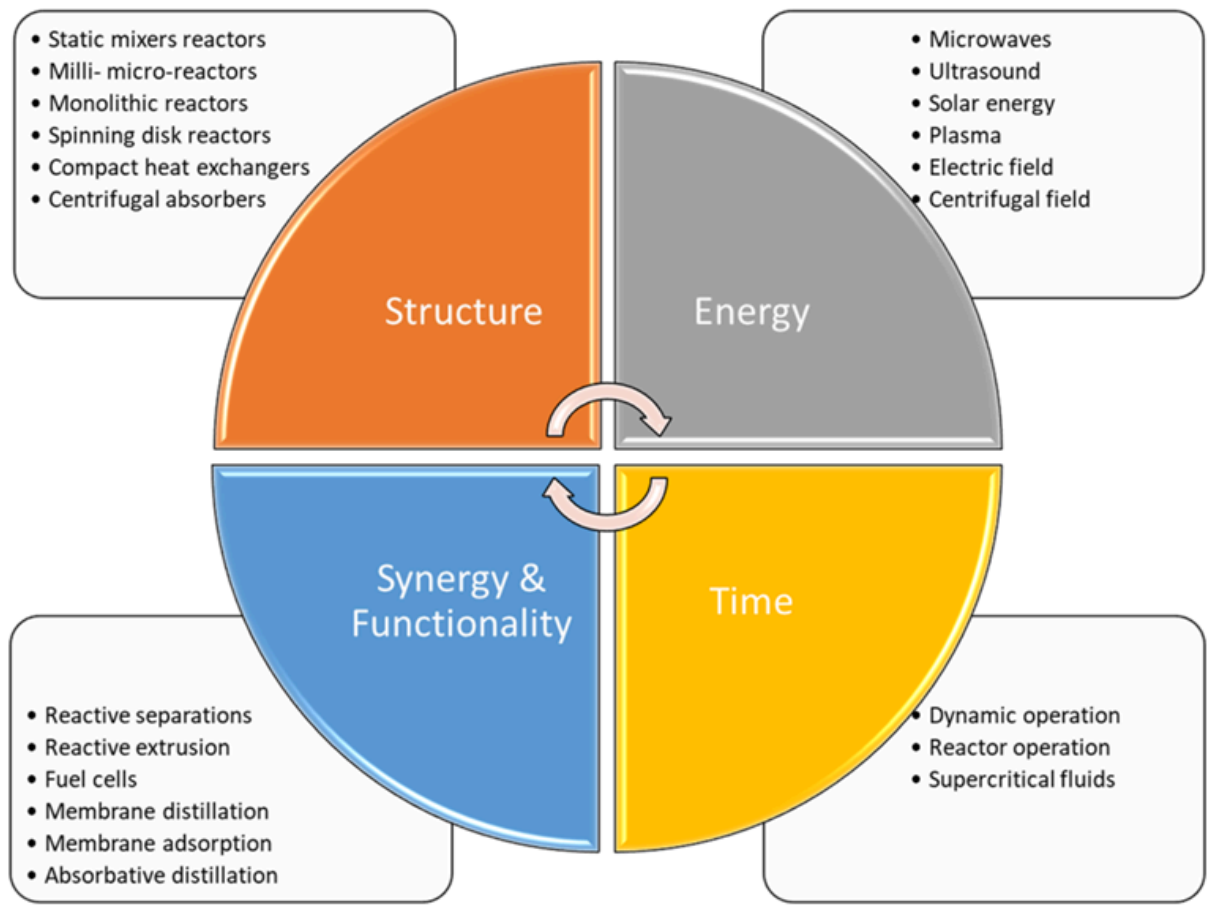

Figure 1. Principles of Process Intensification: definition of the main concepts with application examples (Figure and caption from reference[6] with slight rephrasing under a Creative Commons license 4.0[7])

Without getting into the details of process engineering, simulation, safety, and control, we focus on how advances in chemistry, catalysis, and reaction engineering can further translate to the industrial scale.[5,6,8] Optimizing bench-scale chemistry for high-volume production will benefit greatly from translating laboratory discoveries into commercial success. Artificial intelligence and machine learning (AI/ML) can be key to increasing the efficiency of the reaction engineering process, along with high-throughput experimentation.[9,10] Its impact on chemical reaction engineering will shape the future of process systems engineering, encompassing chemical reactor design, transport phenomena, control systems, and more. Converting bench-scale chemistry into unit operations (such as dissolution, filtration, crystallization, and evaporation) must be demonstrated in efficient process flow diagrams.[11] (Figure 2) The process will mix reactants, purify the products, and either separate the products or recirculate unused reactants. Most of these processes involve control of energy transfer in reactors and transport phenomena. These principles can apply to the chemical industry, petroleum engineering (both upstream and downstream), the polymer industry, paint and adhesive formulations, bioengineering, and even the pharmaceutical industry.[12]

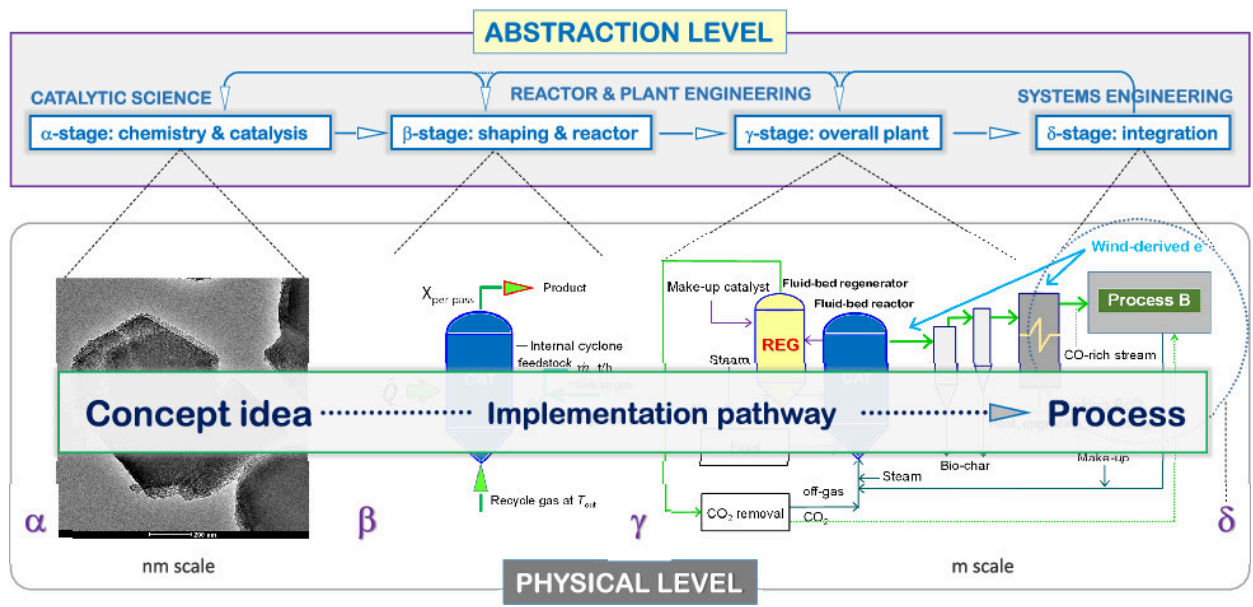

Figure 2.  Stages for developing catalytic materials (Figure from reference[11] with no change under a Creative Commons 4.0 license [7])

## 3.Catalysis:

In chemical reaction engineering, an increase in reaction rate or a decrease in activation energy, $E_a$, is due to the presence of a catalyst (Figure 3).[13]  It is not consumed in the reaction, or at least can be recovered or regenerated, and ultimately reused. Depending on the type of catalyst action, mixing, surface area, concentration, and temperature also affect the reaction rate. Chemical intermediates are isolated in preparation for the next stage. The transition state phenomenon is often the primary focus of the mechanism. The reaction route, or path, is thus determined by the presence or absence of this catalyst, though the product is ultimately obtained.

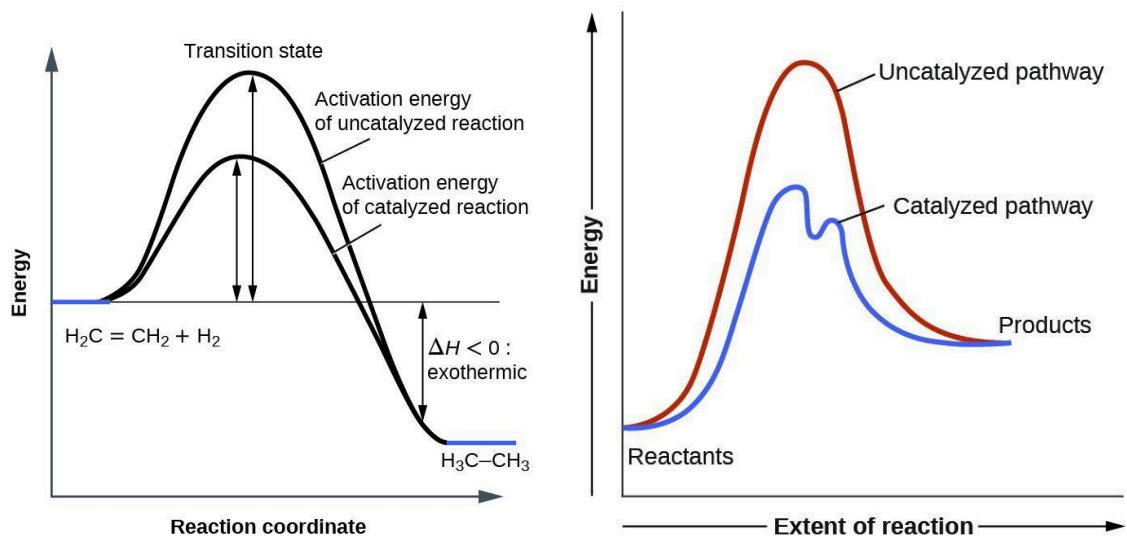

Figure 3. This potential energy diagram shows the effect of a catalyst on the activation energy. The catalyst provides a different reaction path with a lower activation energy. As shown, the catalyzed pathway involves a two-step mechanism (note the presence of two transition states) and an intermediate species (represented by the valley between the two transitions states). (Figures and captions from reference [14] with no change under a creative commons license [7].)

The basic classification of catalysis is homogeneous vs. heterogeneous.[13] The catalytic activity of enzymes and biocatalysts is explained and categorized separately. In homogenous catalysis, the reactants, products, and catalyst are in the same phase. In heterogeneous catalysis, the reactants and catalyst, or even the products, are not in the same phase or separated by a surface or interface. Although it does not change the equilibrium (thermodynamics), it can enable the reaction at lower temperatures, increase the reaction rate, or improve selectivity. Catalysis itself can be classified into pre-catalysis, cooperative catalysis, tandem catalysis, autocatalysis, and switchable catalysis, etc.[13] Enzyme catalysis also has allosteric effects, which can be classified in any of the other categories, including biocatalysts, biological design, and metabolic pathways.[15,16]

In homogeneous catalysis, the catalyst and reactants are mixed in a single phase or in the same solvent, as in proton-catalyzed, organometallic, and field-driven photocatalysis and electrocatalysis.[13] Enzymes or biocatalysts can also be included in the homogeneous realm or in-between heterogeneous states, especially if they are membrane-bound or supported by a porous medium, e.g., enzymes, ribozymes, abzymes, etc.(Figure 4)

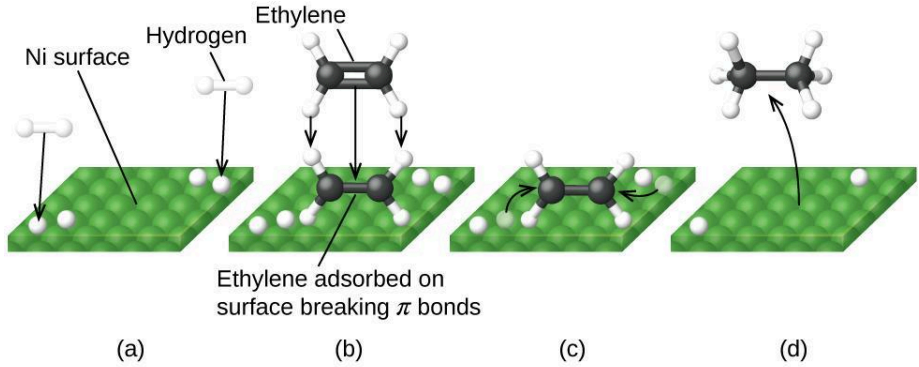

Figure 4. There are four steps in the catalysis of the reaction $C_2H_4+H_2\rightarrow C_2H_6$ by nickel. (a) Hydrogen is adsorbed on the surface, breaking the H–H bonds and forming Ni–H bonds. (b) Ethylene is adsorbed on the surface, breaking the π-bond and forming Ni–C bonds. (c) Atoms diffuse across the surface and form new C–H bonds when they collide. (d) $C_2H_6$ molecules escape from the nickel surface, since they are not strongly attracted to nickel. (Figure and caption from reference[14] with no change under a creative commons license[7])

Heterogeneous catalysis depends on active sites and surface area.[13] Minerals and particles composed of alumina, zeolites, activated carbon, silicon dioxide, titanium dioxide, and others interact with molecules at specific orientations or electronic environments, facilitating bond-breaking and bond-forming.(Figure 5)

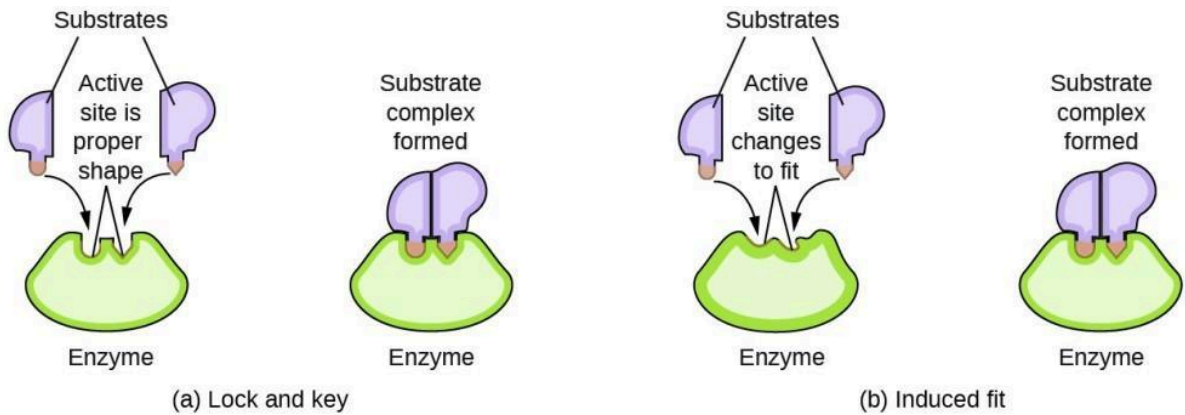

Figure 5. (a) According to the lock-and-key model, the shape of an enzyme's active site is a perfect fit for the substrate. (b) According to the induced fit model, the active site is somewhat flexible, and can change shape in order to bind with the substrate. (Figure and caption from reference[14] with slight rephrasing under a creative commons license[7])

A good summary of heterogeneous catalysts treats them as materials with specific functions and scalability, as they relate to process optimization and systems engineering has been outlined.[11] (Figure 6)  Catalytic materials classifications can include homogeneous, heterogeneous, and enzymatic catalysts – but industrially, the unit price can also include support, co-catalysts, and matrix compositions that are already standard for specific industrial applications, e.g., polyolefins, chemical intermediates production, aromatization, etc.

Other advances in photocatalysis, electrocatalysis, photo-electrocatalysis, mechanochemical catalysis, etc., are worth mentioning, given the interest in field-induced perturbations of energy diagrams and charge transport.[13]

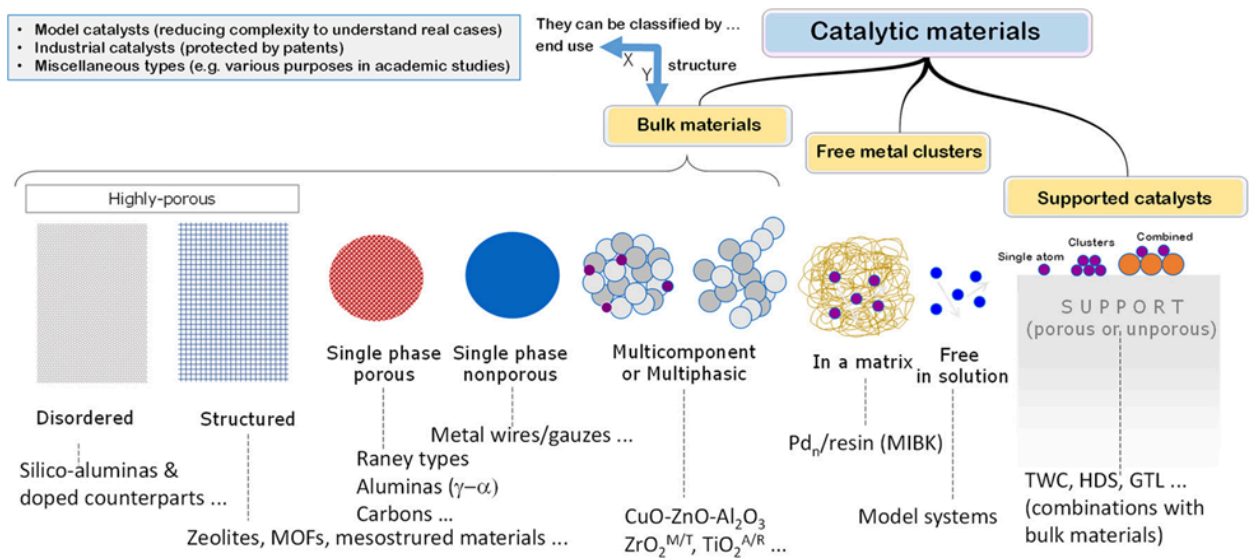

Figure 6. Classification of catalytic materials, in terms of their use ($x$-axis at the top) or structure ($y$-axis, bottom). At the very bottom, a few representative examples are given. Abbreviations: MOFs (metal organic frameworks), $ZrO_{2M}$ (monoclinic zirconia), $ZrO_{2T}$ (tetragonal zirconia), $TiO_{2A}$ (anatase phase), $TiO_{2R}$ (rutile phase), MIBK (methyl isobutyl ketone), TWC (three-way catalyst), HDS (hydrodesulfurization process), and GTL (gas-to-liquids process). Note that supported catalysts can include bulk materials as a support. New concepts, such as single-atom catalysts, would be included in bulk materials or supported catalysts. (Figure and caption from reference[11] with slight rephrasing under a Creative Commons 4.0 license[7])

# 4. Artificial Intelligence (AI) Workflow for Catalyst Design and Optimization

The quests for new catalysts are challenging:[17] 1) several cycles of design, synthesis, and testing are time-consuming, 2) a high number of factors need to be examined in catalytic performance optimization and recovery, and 3) high-throughput screening is needed for new and unaccounted factors in a large-scale and continuous industrial operation. Design of experiment or DOE methods are needed. However, it cannot address the complexity of the catalyst parameter space fast enough. This leaves many opportunities to leverage new discoveries from molecular and quantum-level studies that enable control over transition states and reaction pathways. Other experimental variables in nonlinear chemical reactions are not adequately addressed. Only a Bayesian optimization method can operationalize these discoveries and accelerate catalyst and catalysis research. To bridge the gap mentioned above, many scientists have utilized AI/ML-driven methods to: 1) fine-tune simulation methods, 2) enable effective science literature search and utilize large language model (LLM) transformers, 3) drive the engine of conducting high-throughput and autonomous experiments.[4]

Artificial Intelligence (AI) represents the foundational discipline of engineering systems capable of emulating cognitive functions, such as reasoning, generalization, and problem-solving.[9,10,18] At its core, AI seeks to bridge the gap between deterministic computation and adaptive intelligence observed in biological entities. Within this broad field, Machine Learning (ML) serves as the primary engine of progress, utilizing mathematical algorithms to identify patterns within data. Unlike classical symbolic AI, modern ML paradigms—ranging from data science methods to deep learning—leverage statistical inference to improve performance autonomously as data volume increases, forming the basis for predictive modeling in both industry and academia.

The taxonomy of machine learning is traditionally divided into supervised and unsupervised paradigms, each utilizing distinct algorithmic structures.[10,19–21] Supervised learning involves training models on labeled datasets to map inputs to specific outputs, with Decision Trees and Support Vector Machines (SVM) serving as robust benchmarks for classification tasks. To handle non-linear complexities, these models often employ Kernel Functions, which project data into higher-dimensional spaces to achieve linear separability. In contrast, Unsupervised Learning focuses on discovering latent structures within unlabeled data, primarily through Clustering techniques and Dimensionality Reduction methods such as Principal Component Analysis (PCA), Random Projection, and Multi-dimensional Scaling (MDS), which simplify high-dimensional datasets while preserving their essential variance.

Regression analysis and probabilistic modeling provide the quantitative rigor needed for continuous-variable prediction.[9,10,15,17–19,22] While linear methods offer interpretability, Gaussian Processes (GP) represent a sophisticated Bayesian approach to regression, providing not only predictions but also a formal measure of uncertainty. This probabilistic framework is essential in scientific research where understanding the "confidence" of a model is as critical as the prediction itself. These methods often integrate with Active Learning strategies, in which a model iteratively selects the most informative data points for labeling, thereby optimizing learning in data-constrained environments.

The evolution of Neural Networks has catalyzed a shift toward Deep Learning, where architectures such as Artificial Neural Networks (ANN) and Graph Neural Networks (GNNs) process increasingly abstract representations of data.[23–29] While ANNs are general-purpose approximators, GNNs are specialized for non-Euclidean data structures, such as molecular graphs or social networks. This structural adaptability is particularly evident in the intersection of Natural Language Processing (NLP)[30] and cheminformatics; for instance, the SMILES (Simplified Molecular Input Line Entry System)[31] notation allows chemical structures to be processed as sequences, enabling NLP-based models to perform "molecular translation" and drug discovery tasks with unprecedented efficiency.

Large Language Models (LLMs) represent the current zenith of NLP, utilizing transformer architectures to process and generate human-like text across diverse domains.[30,32,33] These models are trained on petabytes of data to predict tokens in context, resulting in emergent capabilities that transcend simple linguistic mimicry. However, the field is rapidly transitioning from passive generative models to Agentic AI. Unlike standard LLMs that merely respond to prompts, agentic systems are designed for autonomy, utilizing reasoning loops to plan, use external tools, and execute multi-step tasks. This shift marks the move from AI as a chatbot to AI as an autonomous collaborator capable of independent goal pursuit.

To refine these autonomous behaviors, Reinforcement Learning (RL) serves as a critical training paradigm, where agents learn optimal policies through trial and error within a simulated or real-world environment.[34–38] By receiving "rewards" for successful actions, RL agents develop complex strategies that are often non-intuitive to human observers. This paradigm is fundamental to the development of robotics and self-correcting systems. When combined with LLMs' high-level reasoning, RL provides the behavioral framework for Agentic AI to navigate uncertain environments and achieve long-term objectives.

Ultimately, the integration of these diverse methodologies—from classical dimensionality reduction to modern agentic frameworks—signals a move toward more holistic and capable systems. The synergy between structured data representations (like SMILES and GNNs) and the vast reasoning power of LLMs is fostering a new era of computational discovery. As these models become more adept at active learning and autonomous execution, the boundary between data processing and genuine cognitive agency continues to blur, necessitating rigorous academic scrutiny of their ethical and functional limits.

# 5. AI/ML in Catalyst Design and Catalysis: Importance of Data

Advances in the use of AI/ML for chemical synthesis start with design and simulation. Heuristics alone are not enough; lab experience and data-driven approaches are also needed. ML methods will enable the prediction of augmented properties and the formulation of catalyst synthesis steps. What is needed is the ability to apply robust ML models in catalyst design that leverage well-established ML methods to differentiate new data from class-imbalanced data using explainable artificial intelligence (XAI) (Figure 7).[39] However, optimization with ML-driven high-throughput experimentation (HTE) is needed to produce real breakthroughs.[18]

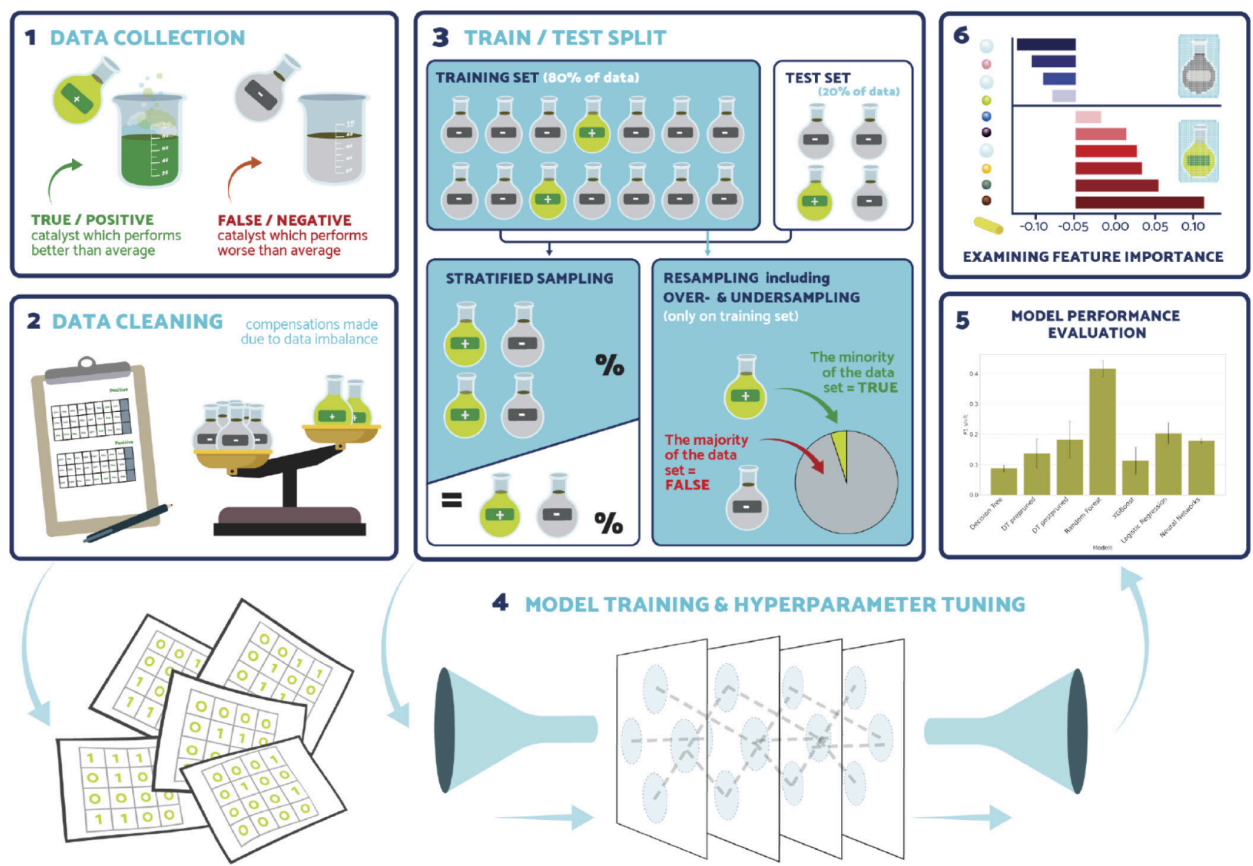

Figure 7. Illustration of the ML framework, starting with data collection and cleaning (steps 1–2), and visualizing the process for obtaining the performance and explanations of a model on a single random train-test split of the data set (steps 3–6). These training and evaluation steps are then repeated for 100 different train-test splits, and the results are aggregated to produce robust performance estimates and feature importance scores. (Figure and caption from reference[39] with no change under a Creative Commons 4.0 license[7])

Beyond synthesis, HTE with catalyst characterization will enable data-driven discovery and, eventually, the optimization of catalyst performance. AI/ML methods will be applied to heterogeneous, homogeneous, and biocatalysis. In heterogeneous catalysis, predicting adsorption energies, understanding orientation dynamics, and elucidating insertion-substitution mechanisms are important.[19,40]

In homogeneous catalyst design, descriptors, together with ML-driven density functional theory (DFT) calculations, will enable high-throughput in silico exploration of the chemical space.[41]

Structure-based enzyme design, based on active sites and substrate-binding pockets, can leverage DL structure prediction methods.[15]

Better utilization of LLMs will give more tools for human-in-the-loop design and testing, and can only accelerate catalyst development. For example, the catalytic language model or CataLM was recently described.[42]

The use of more robust catalyst descriptors can improve the predictive accuracy of ML models and identify key parameters for catalytic activity and selectivity.[43]

New theoretical methods derived from these ML-directed tools and techniques will reveal the structures and reactions in complex catalytic systems, which will eventually be proven by HTEs.[44]

Beyond linear regression and other statistical algorithms, rational catalyst design can be augmented by explainable ML and ML-based candidate prediction and iteration.[45] (Figure 8) The iterative methods can include Shapley Additive exPlanations (SHAP) analysis on specific catalyst systems, Pd metal centers for correlation with higher yields.

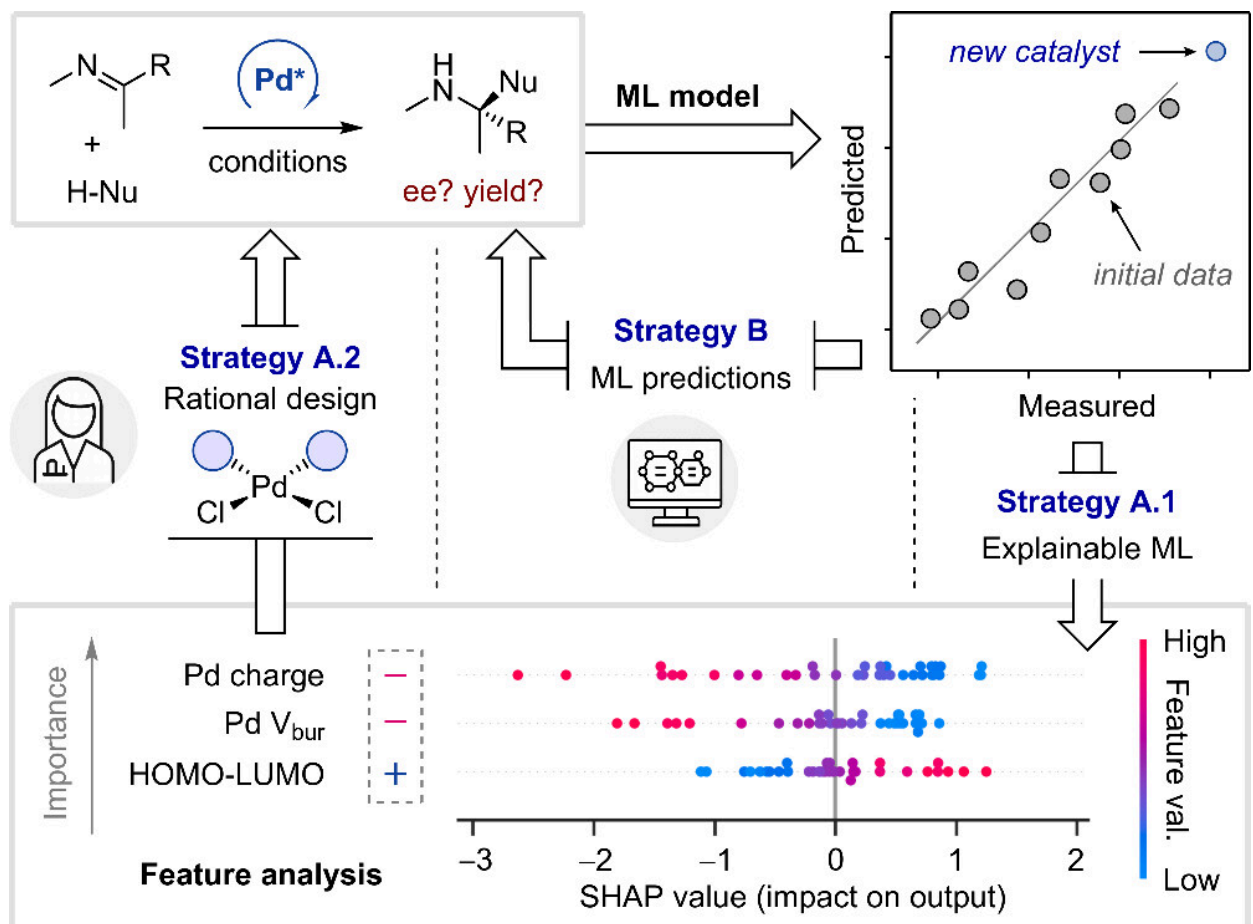

Figure 8. Comparison of different data-driven catalyst discovery strategies: rational design versus ML-based suggestions. (Figure and caption from reference[45] with no change under a Creative Commons 4.0 license[7])

The advances in predictive tools can be elaborated further:
1) The application of DFT, which is computationally expensive, will be accelerated by ML-optimized workflows and algorithms (ML-DFT). The DFT-ML catalyst program, or DMCP was recently employed for optimizing catalytic activity.[46]
2) Graph neural networks (GNNs) were used for new catalyst design protocols. This improved the time for screening, performance prediction, and reaction modeling. This included using various graph-based models to analyze the catalyst's geometry, symmetry, and domain, thereby revealing new critical reaction pathways.[47]
3) The recent advances in dataset availability and ML techniques, together with new generative models, can further enhance new catalyst structure generation. Various autoencoders can generate new hypotheses for crystal diffusion, adsorbate generation, transition-state orientation, and energy and force predictions.

Some other interesting examples of AI/ML-driven catalyst research and development are as follows:

Various alloy compositions were developed for nitrogen activation according to DFT and Bayesian Optiimization,[48] which could be applied to other catalyst research including ammonia synthesis. A number of ML and microkinetic models were implemented to enhance sulfur-resistant catalysts used in steam-methane reforming, which was accomplished through a high-throughput protocol of 500 bimetallic alloy candidates.[49] The identical micro-kinetic modelling and ML optimization were applied to catalysts employed for the CO reduction reaction, resulting in a more efficient computation.[50] AI/ML-driven approaches were deployed to design catalysts for ring-opening polymerization reactions via a domain-specific language known as Chemical Markdown Language (CMDL), achieving predictive models consistent with experimental results.[51] Complex-Solid-Solution (CSS) or high-entropy alloys were simulated for electrocatalyst activities and verified through high-throughput experimentation, with electronic and geometric effects investigated in order to tune catalyst activities.[52] An active learning method was employed to enhance multi-metallic alloy catalysts for hydrogen evolution reactions, focusing on electrocatalytic performance.[53] A review recently tackled the complexity gap of computational heterogeneous catalysis with ML strategies, examining factors related to both mechanistic and chemical structural complexities located at the reactive interface of the catalyst.[54]

Engineering enzymes are gaining importance for chemical conversion as they are considered robust for scale-up.[16,55] There is a growing interest in designing for enzyme stability or predicting catalytic activity via ML-optimized models and database creation.[55] A recent review summarizes AI-based rational design of enzymes with some representative AI algorithms and case studies.[56] Nanozymes are nanomaterials that display enzyme-like biomimicry, while ML techniques were implemented to optimize nanozyme design for substrate selectivity and catalytic activity.[57] An additional notable work outlines AI-assissted protein engineering and design, emphasizing on directed evolution to efficiently establish mutant libraries, involving template-based or template-free tactics.[58] Emerging carrier-free designs for biocatalysts can integrate sustainable hybrid material engineering, incorporating synthetic biology, nanotechnology, and data analytics approaches. Another recent review suggested that this convergence is essential for numerous enzyme catalysis applications.[59] Furthermore, generative AI transforms the reaction-function space of native enzymes into novel functions that serve more industrial applications.[60] This methodology includes the tuning of artificial luciferases, non-heme iron (II)-dependent oxygenases, as well as P450 enzymes towards scale-up with generative AI models.[61]

# 6. AI/ML Workflows with Self-driving Laboratories in Continuous Flow Chemistry

To operationalize HTE with the new catalyst designs and generative AI results for optimum workflows, it is important to realize how continuous flow chemistry (CFC) and microfluidics can operationalize the laboratory. AI in chemistry relies heavily on the generation of good data to be an ultimate tool for scientific discovery.[1] SDLs are becoming the new playground for ML-driven workflows across organic chemistry, pharmaceuticals, nanomaterials, and more. ML together with LLMs is critical for reaction and catalysis engineering to provide good hypothesis-driven objectives for scientific discovery and reaction optimization. SDL is described as the laboratory of the future, a necessary tool to unlock the bottleneck of fundamental science findings and enable translational research.[62–65]

A typical SDL system will consist of components constructed via Python-scripted workflows that integrate with dosing, introduction into the reactor, monitoring, characterization of the products, and the development of a feedback loop mechanism for optimization. An example is the MINERVA framework, which is an autonomous SDL to generate advanced materials that may be adapted for catalyst synthesis.[66] (Figure 9)

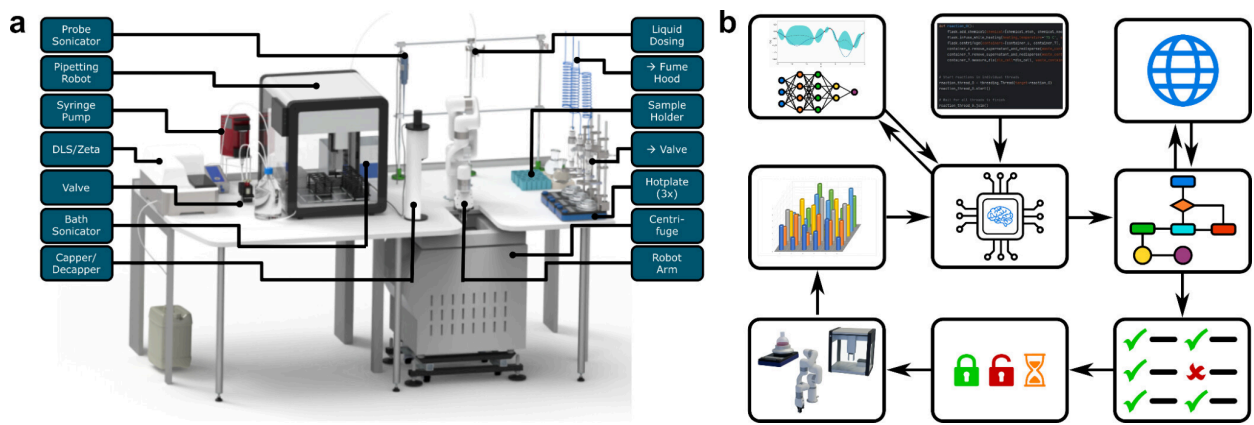

Figure 9. (a) 3D rendering of the SDL platform with the most important hardware components labeled. (b) Schematic representation of a typical workflow. The user supplies high-level synthesis commands as Python code for the API of MINERVA-OS (middle). The backend applies heuristics to determine any values not explicitly specified by the user, potentially looks up missing values for chemicals (such as the density or molar mass) from online resources, performs unit conversions, checks the supplied parameters for integrity and sensibility, determines which hardware will be required in each step, uses the task scheduler to lock/unlock required hardware or enter a waiting queue if the hardware is currently not available, issues the low-level hardware commands to trigger the execution of the high level commands described in the Python script, performs in-line characterization after synthesis and purification of the nanomaterials are complete, and optionally reports the characterization data to a machine learning algorithm to rerun the synthesis with newly suggested, optimized parameters. (Figure and caption from reference[66] with no change under a Creative Commons license 4.0[7])

SDLs enable constrained multi-objective optimization for nanoparticles synthesis and catalyst design.[67] With CFC, the HTE goals can be realized closer to bench-scale chemical engineering and closer to the empirical data generation.[68,69] New reactor designs built on advanced online analytical methods and reliable engineering principles are consolidated by an ML-driven operation.[70] A reported SDL platform[70] can alleviate the knowledge gaps in flow behavior, hydrodynamics, PVFT factorization, multiphase-flow reactions, as well as reactor mixers.[20] Even prototyping with 3D printing for new CFC concepts provides novel reaction and mixing plans, including 3D-printed catalysts.[71] The automation of the CFC for SDL can fill the gap between optimized design and simulation with ML validations. LLMs and agentic AI tasks focus on the problems of reaction engineering and align them with practical process engineering. An SDL operation will deliver standardized data and facilitate automated discovery in chemical reaction engineering.[72,73]

In addition, ML-driven CFC can be paired with reinforcement learning (RL), especially in multi-step processes.[35,74] RL can optimize each step with a feedback loop, training to predict future optimal responses for an autonomous setup. With catalysis and reaction engineering, It will be interesting to apply RL with telescoping reactions in an SDL for each step of the reaction.[75] Advancing to deep reinforcement learning (DRL) will lead to self-optimization strategies for discovery science in reaction engineering[34], and hyperparameterization. Libraries and datasets for LLMS will be quickly built up[76] for more scalable synthesis chemistry and discovery.[77]

CFCs and SDL can also be further applied to microfluidics and millifluidics for scale-down and miniaturization,[78] including online monitoring as an HTE discovery platform.[79] It was actually the other way around when most of the important organic reactions and polymerization optimizations were first demonstrated in microfluidics. ML-driven microfluidics will enable more predictive reaction control at much smaller volumes, suitable for handling expensive or rare catalysts or reagents.[80] An SDL in a miniaturized setup can help evaluate many new concepts in reaction engineering and the synthesis of nanomaterials in flow.[20,81,82]

Nanoparticle synthesis is particularly interesting for ML-optimized synthesis and holds great promise for novel catalyst designs due to its high efficiency and surface-to-volume ratio.[66,83–85]

In polymer reaction engineering, an SDL-based polymer synthesis can provide important control over molecular weight, sequence polymerization, microstructure, branching, topological control, macrocycles, etc.[86] This includes optimization of polymerization methods, including emulsion polymerization.[87] We have reported on the use of SDLs to answer questions in ML-Driven copolymerization and demonstration of feedback loop protocols.[21,88]

In summary, an ML-enabled CFC platform is crucial for rapid, adaptable optimization of multiple reaction metrics,[89] and swift screening in catalysis and reaction engineering.[90] An SDL setup guided by RL and BO becomes a powerful tool for self-optimization in autonomous discovery.[35,91]

# 7. Specific examples of Catalysis, AI/ML, and SDL together

In "digital catalysis," the design space is now wide open. With AI-driven high-throughput synthesis, it is feasible to obtain bigger experimental datasets with swifter feedback-loop control.[92] SDLs enable automated systems and intelligent algorithms. The data-in-the-loop also needs to be closed with the human-in-the-loop operation. The SDL will not be a fully autonomous station. When knowledge is acquired from a chemical plant or adaptive experiments, the setup can be more effective when used by an expert in an active loop process.[93] An agentic AI and task-driven role can still be more efficient in the hands and eyes of the expert scientist. The challenges of an autonomous catalyst process require an all-hands-on-deck approach.[94]

Previous works have highlighted the use of AI/ML-optimized reaction engineering and HTE for the oxidative coupling of methane, ethanol-to-butadiene conversions, $CO_2$ hydrogenation, as well as hydroformylation reactions.[95,96] An SDL system "Fast-Cat" for rapid catalyst development has been reported.[97] An autonomous Pareto-front visualization was implemented for homogeneous catalyst development in high-temperature, high-pressure gas–liquid reactions, implemented to hydroformylation reactions of syngas and olefins.[98] An enhanced ML-driven enzymatic reaction intensification was demonstrated in an SDL platform.[99] Copilot for Real-world Experimental Scientists (CRESt) enabled the discovery of multimetallic catalysts with a large vision–language multidimensional model and BO in one SDL.[100] These concepts were outlined in a prospective article for advancing catalysis research by incorporating autonomous and digital catalysis as a true human–AI–robot collaboration.[62] This further validates a roadmap for transforming catalysis with AI/ML tools in an SDL setting.[64,101]

Furthermore, generative and agentic AI can automate the generation of models.[28,29,61,100,102] Quantifying uncertainty and coupling theory with experiment can provide an interpretable, and transferable application for real-world catalytic systems.[103] Recently, a more flexible and affordable SDL for automated reaction optimization, called RoboChem-Flex, has been reported.[104]

# 8. Conclusions and Future Perspectives.

To enable advances in AI/ML for chemical synthesis and reaction engineering, a strong focus on catalyst design and testing is a significant step forward. This starts with a robust workflow on ML-driven design and simulation. This has gained ground in the catalysis community, where many leads and theoretical frameworks have already been established. However, integrating this in the lab experience and data-driven approaches can only accelerate this development. HTEs with SDL will need to achieve real breakthroughs that can enter the pilot scale to chemical plant processes. Beyond just synthesizing the catalyst, an autonomous HTE with catalyst characterization will enable true parametrization and create a virtuous circle for data-driven discovery. AI/ML methods will be applied to heterogeneous, homogeneous, and biocatalysts to predict economic viability and energy efficiency in the chemical process industries.

**Funding and Acknowledgement**

This work was supported by the Center for Nanophase Materials Sciences

(CNMS), which is a US Department of Energy, Office of Science User Facility at Oak Ridge National Laboratory, and Laboratory Directed R&D (ORNL INTERSECT).

## Conflicts of Interest and Declaration

The author declares no conflict of interest. Also, an archived version of the manuscript was submitted as a non-peer-reviewed preprint in an archiving portal.

## Author Contributions

RCA planned the layout. RCA and JC all contributed to the overall design, literature review, revision, and writing.

**Last update: 2.25.2026**